\documentclass[doublecol]{epl2}
\usepackage{amsmath}
 
\newcommand{\ud}{{\mathrm d}}

\title{The dynamics of the DNA denaturation transition}
\author{Titus S. van Erp$^1$ and Michel Peyrard$^2$} 
\institute{
\inst{1} Centrum voor Oppervlaktechemie
en Katalyse, KU Leuven, Kasteelpark Arenberg 23, B-3001 Leuven,
Belgium\\
\inst{2}
Laboratoire de Physique, Ecole Normale Sup\'erieure de Lyon,
46 all\'ee d'Italie, 69364 Lyon Cedex 07, France
}
\abstract{
The dynamics of the DNA denaturation is studied using the Peyrard-Bishop-Dauxois model.  The denaturation rate of double stranded polymers decreases exponentially as function of length below the denaturation temperature. Above $T_c$, the rate shows a minimum, but then increases as function of length. We also examine the influence of sequence and solvent friction.  Molecules having the same number of weak and strong base-pairs can have significantly different opening rates depending on the order of base-pairs.
}
\begin{document}
\maketitle
For decades, experimental and theoretical scientists have been fascinated by
the thermal DNA denaturation~\cite{DNAden}. It is biologically relevant since the 
opening of the double helix in an important step for the
transcription of the genetic code~\cite{trans} but it is
  also becoming important for nanotechnology as DNA is now used for
  its self-assembly properties \cite{Goodman}, to create nanodevices
  \cite{Komiya}  or to design molecular memories \cite{Takinoue}.
The different AT and GC base-pairs (bps) are intra-connected by two and
three hydrogen-bonds, respectively. Denaturation experiments, in which 
UV absorbance is measured as function of temperature, show  
discrete steps associated to the sequential opening of soft and
stiff regions in the DNA sequence. These regions mainly differ by their 
density of weak AT- versus strong GC-bps, but also the specific order 
is important~\cite{Dornberger}.
On the other hand,
large homogeneous chains denature in a single step within a very small 
temperature interval,  a remarkable realization of  an effectively 
one-dimensional phase transition~\cite{Inman}.

Theoretically, the statistics of DNA have been modeled 
via Ising type models~\cite{poland} 
in which each bp is given a value 0 or 1 depending on whether it is 
open or closed.
This approach allows the study of 
extremely large sequences but cannot be used to study 
dynamics. This  has been the 
principle motivation for Peyrard, Bishop and Dauxois 
to develop a continuous model (PBD)~\cite{PBD}.
The PBD model is computationally somewhat more expensive than the Ising type models, but
still quite efficient due to its mesoscopic character. As a result, the PBD model is very well suited to 
study dynamics of reasonably long DNA molecules at timescales that are not reachable  with full-atom simulations~\cite{fullatom}.
The PBD model, therefore, meets our requirements that we need to study the dynamics of the DNA denaturation transition. In specific, 
we want to understand how the rate of full denaturation depends on several parameters such as length, sequence, solvent friction, and temperature.
 
The PBD model has revealed a rich spectrum of dynamical phenomena
such a the existence of nonlinear localized modes~\cite{breather1}.
Ironically, most studies on the PBD model that apply 
specific types of dynamics, such as Nose-Hoover, Brownian motion, and Langevin 
(sometimes even using 
a configurational dependent friction parameter~\cite{LangePBD}), have reported 
on equilibrium properties like the fraction of open bp's at a given 
temperature. These properties, in principle, do not depend on the 
dynamics~\cite{commentDas}.
So far, the study on actual dynamics has been limited to the local opening
of DNA~\cite{bublifetime}
or denaturation at elevated temperatures~\cite{rouse}.
For applications such as nanodevices \cite{Komiya} on
  molecular memories \cite{Takinoue}, the timing becomes important.

The dynamics of full denaturation is difficult to access 
using standard MD as it is a rare event on the timescale that is achievable
by molecular simulations. Transition interface 
sampling (TIS)~\cite{ErpMoBol2003} and the more recent Replica Exchange 
TIS~\cite{vanErp07PRL} are powerful techniques to beat this timescale problem.
However, a very efficient implementation
of the reactive flux (RF) method~\cite{FrenkelSmit} can treat the PBD model even more efficiently~\cite{vanErp07PRL}.
The RF method starts from the Transition State Theory
(TST)  expression, but corrects for correlated recrossings via 
a dynamical factor, the transmission coefficient. Unlike TST, the RF method provides an exact expression for the rate constant that is independent 
to the choice of reaction coordinate (RC).
In Ref.~\cite{vanErp07PRL}, we showed how the free energy and the transmission coefficient can be calculated very efficiently
by  exploiting specific peculiarities of the PBD model. 
Using this approach, 
we can achieve rates far below  $10^{-15}$ ns$^{-1}$ or, in other words, relaxation times of several days or even years.
We are not aware of any mesoscopic modeling approach that is able to reach such timescales.
In this letter, we apply this method to determine the dependence
of the denaturation rate of double stranded DNA as function of its length,
the temperature and solvent friction, the content of weak AT versus strong CG bp's, and their specific order.

The PBD
describes the DNA molecule
as an one-dimensional chain of effective atom compounds
yielding the relative base-pair
separations $y_i$ from the ground state positions.
The total potential energy $U$ for an $N$ base-pair DNA chain is  given by
$U(\{y_i\})=V_1(y_1)+\sum_{i=2}^N V_i(y_i) +  W(y_i,y_{i-1})$
with 
\begin{align}
V_i(y_i) &=D_i \Big( e^{-a_i y_i}-1\Big)^2  \label{eqPBD} \\
W(y_i,y_{i-1}) &= \frac{1}{2} K \Big( 1+\rho e^{-\alpha(y_i+y_{i-1})}\Big)(y_i
- y_{i-1})^2 \nonumber
\end{align}
The first term $V_i$ is the Morse potential describing the
hydrogen bond interaction between bases on opposite strands.
$D_i$ and $a_i$ determine the depth and width of this potential
for the  AT and GC base-pairs. Note that the PBD model does 
not distinguish between  A- and T-  nor between G- and C-bases. 
The second term $W$ is the stacking 
interaction.  
The $\rho$-term   
makes that the effective strength of the stacking interaction drops from $K(1+\rho)$ down to $K$ whenever either $y_i$ or $y_{i-1}$ becomes significantly 
larger than $1/\alpha$.  
This effect mimics the decrease of overlap between $\pi$- electrons when one of two neighboring bases move out of stack and it is thanks to this that the sharp phase transition 
in long homogeneous chains can be reproduced.

Our results are primary focused on the parameter 
set of Campa and Giansanti~\cite{CAGI} with
$K=0.025$ eV/\AA$^2$, $\rho=2$,
$\alpha=0.35$~\AA$^{-1}$, $D_{AT}=0.05$ eV, $D_{GC}=0.075$ eV,
$a_{AT}=4.2$~\AA$^{-1}$, $a_{GC}=6.9$~\AA$^{-1}$.
However, we will also shortly investigate the more recent parameter set by 
Theodorakopoulos~\cite{Theod}: 
$K=0.00045$ eV/\AA$^2$, $\rho=50$,
$\alpha=0.2$~\AA$^{-1}$, $D_{AT}=0.1255$ eV, $D_{GC}=0.1655$ eV,
$a_{AT}=4.2$~\AA$^{-1}$, $a_{GC}=6.9$~\AA$^{-1}$. 
Particular of this data-set is the very high $\rho$ value which 
should reflect the large difference in persistence length of 
single stranded and double stranded 
DNA~\cite{persis}.
$\lambda\equiv \min[\{ y_i \}]$  was chosen as 
reaction coordinate
RC~\cite{vanErp07PRL} and $y_0=1$ \AA ~as the opening threshold. Henceforth,  $y_i>y_0$ 
implies that base-pair $i$ is open and $\lambda>y_0$  that 
the complete molecule is denatured.

The RF  method expresses the overall reaction rate
as an equilibrium probability density to be at a surface on the barrier 
(here defined by $\lambda(\{ y_i \})=y_0 )$, under the condition that the system is at the reactant side of this surface,
times
a dynamical transmission factor.
 \begin{align}
k=P(\lambda=y_0|\lambda \leq y_0) \times R \textrm{ with }
R=\left \langle 
\dot{\lambda} \theta(\dot{\lambda})  h^b_{0,y_0}
\right \rangle_{\lambda=y_0}
\label{RF}
 \end{align}

\begin{figure}[ht!]
  \begin{center}
  \includegraphics[angle=-90,width=6.6cm]{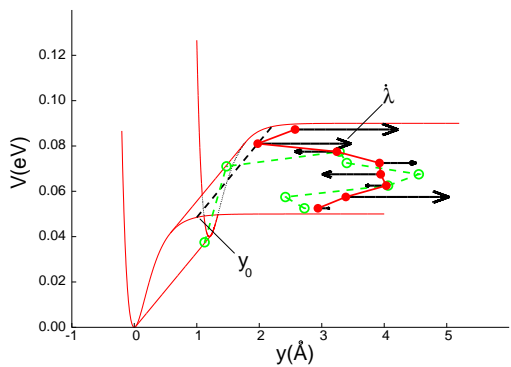}\\
   \caption{(color online)  The above figure illustrates an example  configuration (solid red spheres) at the surface $\lambda(\{y_i\})=y_0$ that was generated by our algorithm
   for a homogeneous $N=8$ AT-chain. It has one particle that is exactly at $y_0$ while $y_i>y_0$ for all others.  
   The arrows envision the corresponding Maxwellian velocities. 
   Each of these sampled phases points will give a single value $x$ that is either zero or positive. If $\dot{\lambda}$ (which is simply the velocity of the particle that is exactly on the surface) equals less than zero, then $x=0$. If $\dot{\lambda}>0$, all velocities of the system will be reversed and the dynamics are propagated using the Langevin
   dynamics. These dynamics are continued until one particle reaches the well ($y_i < 0$ for one particle or  $\lambda<0$) or until all particles move beyond the initial surface 
   ($y_i>y_0$ for all particles or $\lambda>y_0$). In the last case $x$ is assigned zero again. In the first case $x$ equals the initial velocity  $\dot{\lambda}$.
   The final configuration is shown by the green open circles. A particle at the end of the chain has moved inside the well of the Morse potential (hence $x=\dot{\lambda}$ for this case).  This event implies a commitment to the natured state; this particle will quickly pull the others inside the well and a rapid recrossing with the $y_0$ surface is highly unlikely after this point.
    The unnormalized transmission coefficient is the average of the sampled values $x$: $R=\bar{x}$.
\label{figx}}
  \end{center}
\end{figure}

The dynamical factor $R$ is called is the unnormalized transmission coefficient.
The brackets 
with subscript
denote an ensemble average
that is constrained on
the surface $\lambda(\{y_i\})=y_0$.
It is calculated by releasing 
 dynamical trajectories forward and backward in time starting from a proper equilibrium 
ensemble of configuration points on
this surface
and  Maxwell-Boltzmann distributed velocities. 
$h^b_{0,y_0}$ is a binary function 
that is 1 if the backward trajectory from such a point evolves 
to the minimum of
the Morse potential ($\lambda=0$ \AA) without recrossing 
the initial surface ($\lambda=y_0$). 
Otherwise it is 0. 
Since this surface is at the frontier of what is considered to be the product state,
no forward trajectory is needed as in Ref.~\cite{van06}, since $\lambda(\{y_i\})=y_0$ and
$\dot{\lambda}>0$ 
implies that the product state will be entered within an infinitesimal small time step.
This procedure is sketched in fig.~\ref{figx}.

Assuming $h^b_{0,y_0}=1$ whenever $\dot{\lambda}>0$ would result into the TST approximation of the reaction rate.  
In practice, $h^b_{0,y_0}=0$ for the vast majority of trajectories though its average value remains measurable for this RC due to the flat plateau of the Morse potential.  
The time duration of these trajectories are much shorter than the actual relaxation time that relates to the long time evolution of the system having 
relatively  small oscillations.
Therefore, even though the very large fluctuations of these short trajectories
are at the limit of what the PBD can describe accurately, their short duration, with respect to the actual relaxation time,
make quantitative rate calculations reliable. Also the fact that we stop our trajectories whenever $y_i>y_0$ for all $i$ is reasonable.
Besides that the PBD model isn't too accurate beyond this point, it is expected that the chance of reclosing is even smaller in a more accurate three-dimensional model.
We can, therefore, assume that the system is committed to the denatured state beyond this point and it is therefore sufficient to stop the trajectory
whenever the $y_0$ transition surface is crossed.
Still, it is fair to say that our approach and RC will probably not work for more elaborated potentials that contain explicit   potential barriers for reclosing~\cite{closingbarrier} since
it will result in $R \ll 1$.

The equilibrium density is related to the free energy via $F(\lambda)=
-k_B T \ln P(\lambda)$ with $k_B$ the Boltzmann constant and $T$ the
temperature in Kelvin.  The surface does not necessarily have to be at
the local maximum of the free energy profile, the transition state (TS),
though this is generally the most efficient choice since it maximizes
the transmission $R$.  In our case the surface $\lambda(\{y_i\})=y_0$ is
slightly beyond the TS at the beginning of the denaturation state or
product state region. This has, however, no effect on the final results since
the transmission coefficient  and probability density 
are like communicating vessels; a too high value for $P$ (or too low value of $\Delta F$) is compensated 
by a lower transmission $R$. Therefore, the calculated rate values are insensitive to the exact location of the transmission surface.

Because of the first neighbor character of the PBD model, the free energy or probability density $P(\lambda=y_0|\lambda\leq y_0)$
can be computed very efficiently by means of an iterative numerical integration scheme~\cite{vanErp07PRL}.
\begin{align} 
P
=
\frac{\sum_i \int \ud y_1 \ldots \ud y_N  \delta(y_i-y_0)
\prod_{j \neq i} \theta(y_j - y_0)
e^{-\beta U(\{y_i\})}}{\int \ud y_1 \ldots \ud y_N 
(1-\prod_k \theta(y_k-y_0) ) e^{-\beta U(\{y_i\})}} 
 \label{PA}
\end{align}
Now, as the integrals of Eq.~(\ref{PA})  are all of a special factorial 
form~\cite{VanErpPRL}, we can apply the direct integration
method of~Ref.\cite{VanErpPRL} using an  
integration step of $dy=0.05$~\AA~ and integration boundaries
$|y_i-y_{i-1}|<d=\sqrt{2 |\ln \tau|/\beta K}$ and 
$
-1/a_{AT}\ln \Big[ \
\sqrt{ |\ln \tau|/\beta D_{AT} }+1 \Big]<
y_i < y_0+\sqrt{N} d$. The tolerance $\tau$ is set to $10^{-40}$ 
which implies that all contributions of
$e^{-\beta V(\{y_i\})}$ lower than this value are neglected.
The very low value of $K$ in the parameter set
of ~\cite{Theod}
with respect 
to~\cite{CAGI} 
implies significant larger integration boundaries and a 50 times higher
computational cost. 

In the next step, we need to generate a representative set of configurations
on the surface $\lambda(\{y_i\})=y_0$. Also this can be achieved very
effectively for this particular model and RC.
We make use of the fact that ensemble constraint to this surface 
is identical to a
freely moving chain on a translational invariant potential $U'$ with 
$U'(\{y_i\}) \equiv U( \{y_i-\lambda(\{y_i\})+y_0\})$~\cite{VanErpPRL}.
Therefore, 
we generate the required surface points by running a MD simulation using 
potential $U'$, save every 1000th time step to dissolve correlations. Then, we 
shift 
these configurations back to the surface $\lambda(\{y_i\})=y_0$. From each 
point, we
release a backward trajectory using normal potential $U$ and 
determine $\dot{\lambda} \theta(\dot{\lambda})  h^b_{0,y_0}$. 
We 
averaged over $10^6$ trajectories using a time step of 
1 fs and bp masses of 300 amu. The temperature was controlled using Langevin
dynamics with a friction coefficient of $\gamma=50$ ps$^{-1}$ unless 
stated otherwise.

It is important to stress that factor $P$ is purely a thermodynamic
property (i. e. independent of $\gamma$ or masses) while $R$ depends on the full dynamics.
This methodology allows us for the first time to examine the theoretical PBD denaturation rates  
of long DNA molecules
as function of sequence, 
solvent friction, temperature, and model parameters.

\begin{figure}[ht!]
  \begin{center}
  \includegraphics[angle=-0,width=6.6cm]{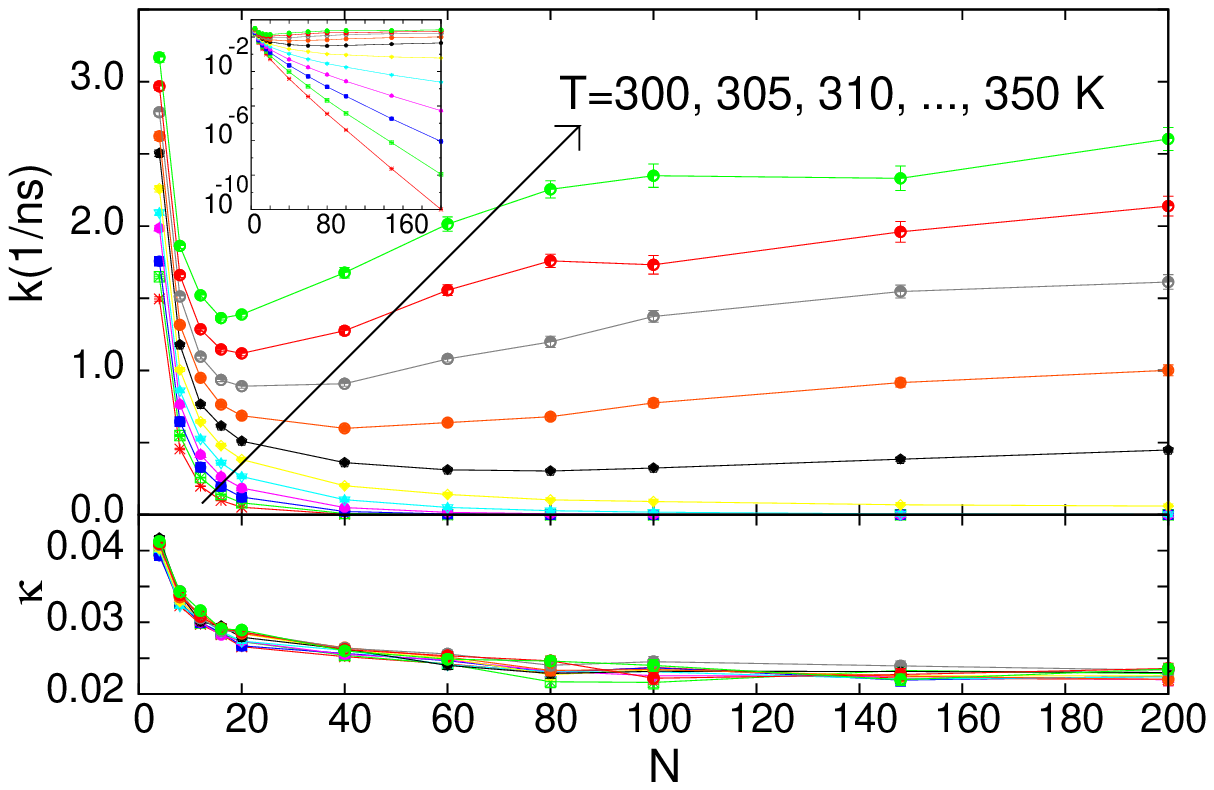}\\
   \caption{(color online) Top: Denaturation rate of homogeneous AT chains as function of the number of bps at different temperatures. Inset shows the same results in log-scale. Bottom: The corresponding transmission coefficients.
\label{figkvsN}}
  \end{center}
\end{figure}

Fig.~\ref{figkvsN} 
shows the calculated denaturation rates of homogeneous A/T-DNA sequences of different lengths and temperatures. Below the critical denaturation temperature ($T_c \approx 325 K$) 
the denaturation rate of the DNA molecule is exponentially decreasing as function of its length. Above $T_c$, there is an initial decrease as function of the 
number of bps $N$, but then starts to increase again. 
Both the depth and the position of the minimum decrease as
function of temperature. This feature  is mainly an effect of  
equilibrium statistical physics (factor $P$) rather then dynamics 
(factor $R$) as we can conclude from
the lower panel of fig.~\ref{figkvsN}.  This panel shows the normalized transmission 
coefficient $\kappa \equiv R/\left \langle \dot{\lambda} \theta(  \dot{\lambda}  )  \right \rangle=\sqrt{2 \pi \beta m} R$ 
which is equivalent to the true rate constant divided  by the transition state theory (TST) value. 
The results show that, although 
$\kappa$ is far smaller than one,
it is more or less constant with respect of temperature and chain length. Only for 
short polymers $N<20$ there is a noticeable upturn of $\kappa$.   Henceforth, despite that TST
overestimates  the rate by more than a factor 40, the relative TST rates are approximately correct.

An increase of denaturation rate as function of its length is remarkable even for $T>T_c$. 
For macroscopic systems having two phases $A$ and $B$, crossing the phase transition temperature coincides 
with a discontinuous jump of the equilibrium constant from zero to infinity. 
Since the equilibrium constant is related to the ratio of forward and backward rate constants, 
$k_{A\rightarrow B}/k_{B\rightarrow A}$  goes from zero to infinity in the limit
 $N\rightarrow \infty$
when $T$ crosses $T_c$.   
Still, this does not imply that  the absolute values of  any of the two rates should increase as function of $N$. In fact, 
the decreasing denaturation time $\tau \sim 1/k$ as function of $N$ is in contrast to the 
power-law scaling
$\tau \propto N^{2.57}$ of a non-equilibrium case study~\cite{rouse}.

\begin{figure}[ht!]
  \begin{center}
  \includegraphics[angle=-0,width=6.6cm]{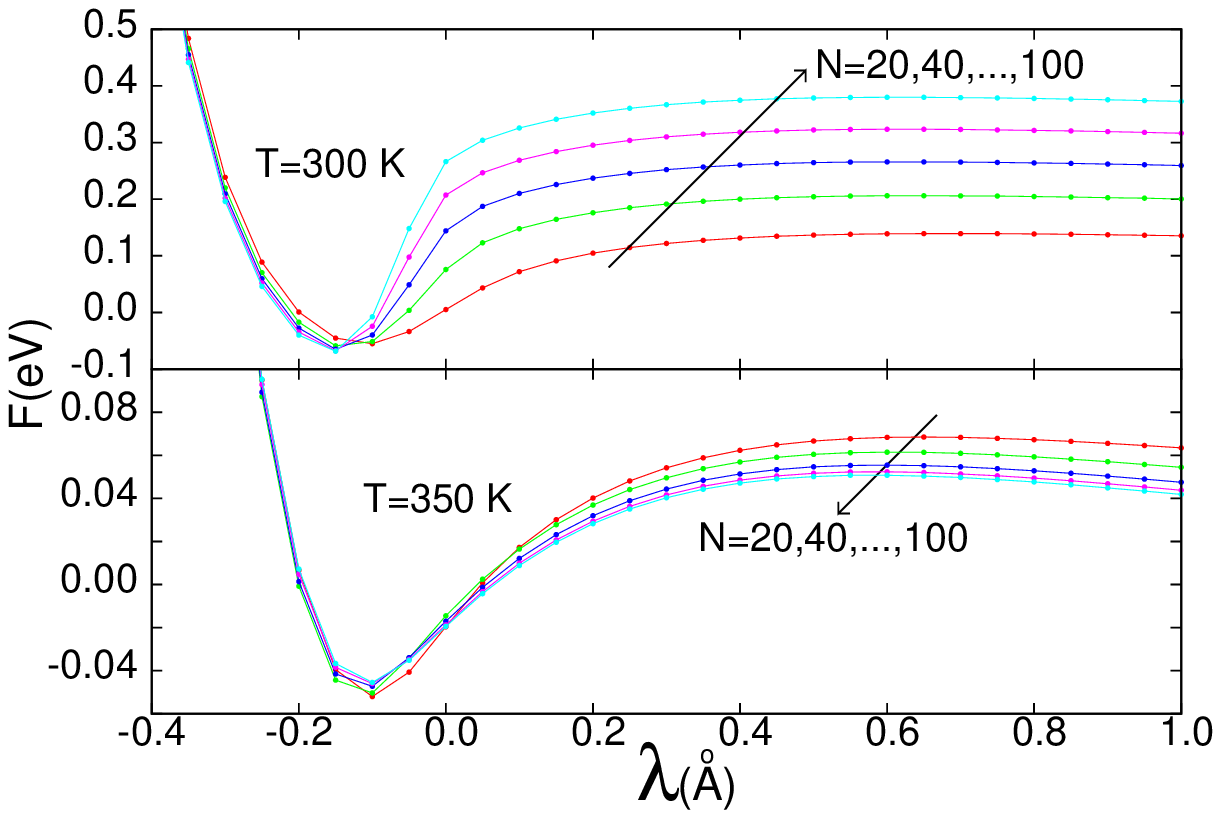}\\
   \caption{(color online) The free energy profile as function of the RC for homogeneous A/T chains of different lengths and temperatures $T=$300~K (top) and $T=$350 K (bottom).
\label{figFE}}
  \end{center}
\end{figure}

To examine the effect of temperature on the equilibrium statistics, 
fig.~\ref{figFE} 
shows the
free energy $F(\lambda')=-\ln \left [ P(\lambda=\lambda'|\lambda<y_0)/
P^0 \right]/
\beta$ with $P^0=1/$\AA ~being an arbitrary reference density.
At temperature $T=300$ K the free energy barrier is clearly increasing as function of chain length while it is slightly decreasing at $T=350$ K.
The longer polymers profit from the larger entropy gain at high $T$.

\begin{figure}[ht!]
  \begin{center}
  \includegraphics[angle=-0,width=6.7cm]{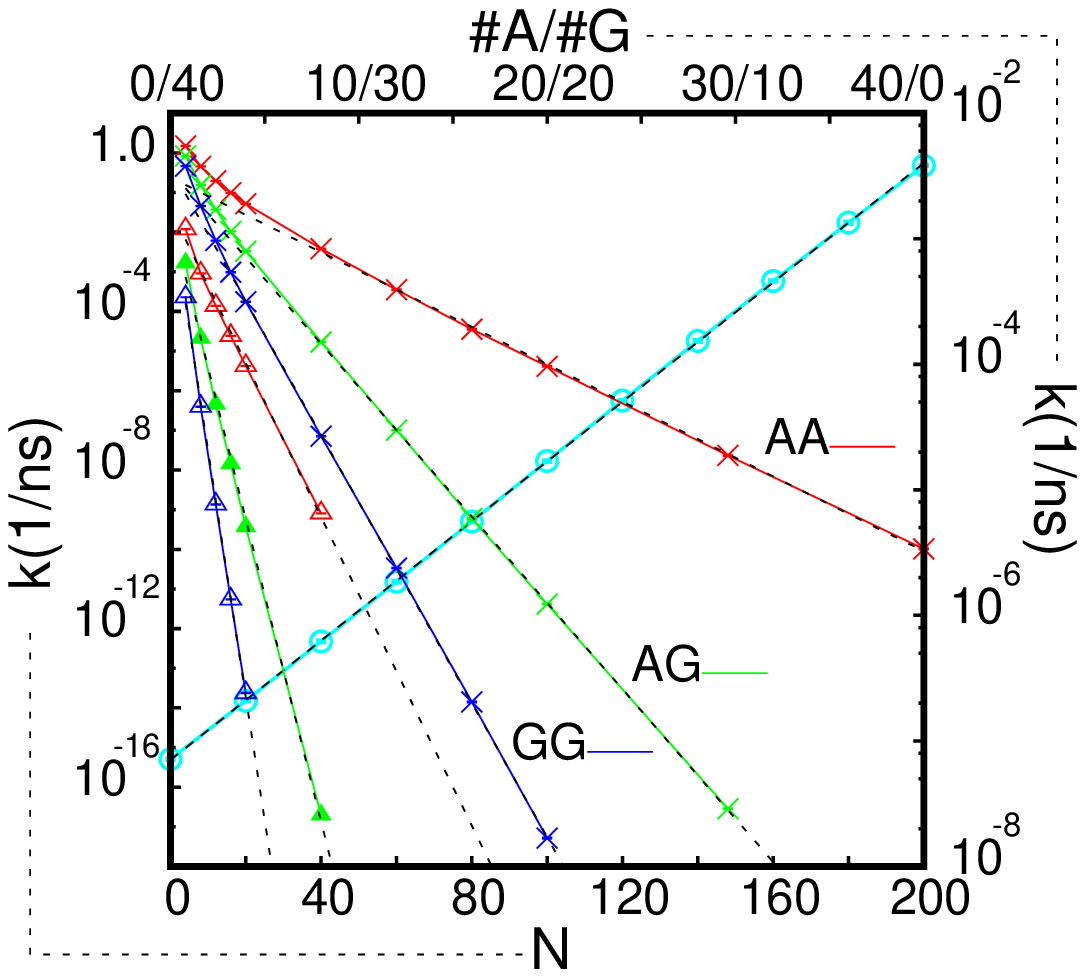}\\
   \caption{(color online) Denaturation rate versus the number bps 
for homogeneous A/T- (red line) and G/C-chains (blue) and a 
sequence having $N/2$ consecutive A's followed by $N/2$ consecutive 
G's (green).
Crosses correspond to the parameter set of Ref.~\cite{CAGI}, triangles
correspond to Ref.~\cite{Theod}. Blue line 
indicates the rate (right and top axis) for a mixed 40 bp chain having $M$ consecutive A- followed by 40-M G-bps. Dashed black lines are best linear fits in the log-plot. 
\label{figkvsN2}}
  \end{center}
\end{figure}

In fig.~\ref{figkvsN2},
we examined the effect of heterogeneities in the DNA sequences at room 
temperature conditions ($T=300 K$). The red and blue curve
represent the denaturation rate for the homogeneous A/T and G/C-sequences.
The green curve in the middle corresponds to sequences
where the first half is purely A/T and the other half  G/C.
All curves become straight lines for 
sufficiently large $N$ which implies an exponential dependence on the chain length.
Linear fits in the log-plot reveal that $k \approx a b^N$ with
$a(1/{\rm ns}),b=0.240, 0.897$ for the A/T- and 0.436, 0.678 for the G/C-chain.
The mixed AG-sequence has $a=0.346$ and $b=0.778$. The 
latter value is very close to $\sqrt{b(A)\times b(G)}=0.780$
which suggest $k \propto b(A)^{N_A} 
b(G)^{N_G}$  as a general rule for a mixed sequence where $N_A$ and $N_G$ are the number of A/T- or G/C-bps. The results based on the alternative 
parameter set of Ref.~\cite{Theod} show the same trend but predicts 
denaturation rates that are orders of magnitude lower. The linear fits
give $a,b=0.039, 0.639$ for A/T, 0.048, 0.240 for GC-, and
0.0249, 0.415 for the mixed chain. The latter is again very close
to $\sqrt{b(A)\times b(G)}$ which corroborates with the above.
In fig.~\ref{figkvsN2} we also show the denaturation rate
for a polymer of fixed $N=40$ length as function of the A,G-content
using parameter set~\cite{CAGI}.
Like before, the order of the polymer is such that the A- and G-bps are placed
consecutively at opposite sides. Making the linear fit in the log-plot shows 
that $k\approx 7.06 (1/$ns$) 1.31^{N_A}$ which is again in excellent agreement
with the above rule since $b(A)/b(G)=1.32$.

\begin{figure}[ht!]
  \begin{center}
  \includegraphics[angle=-0,width=5.0cm]{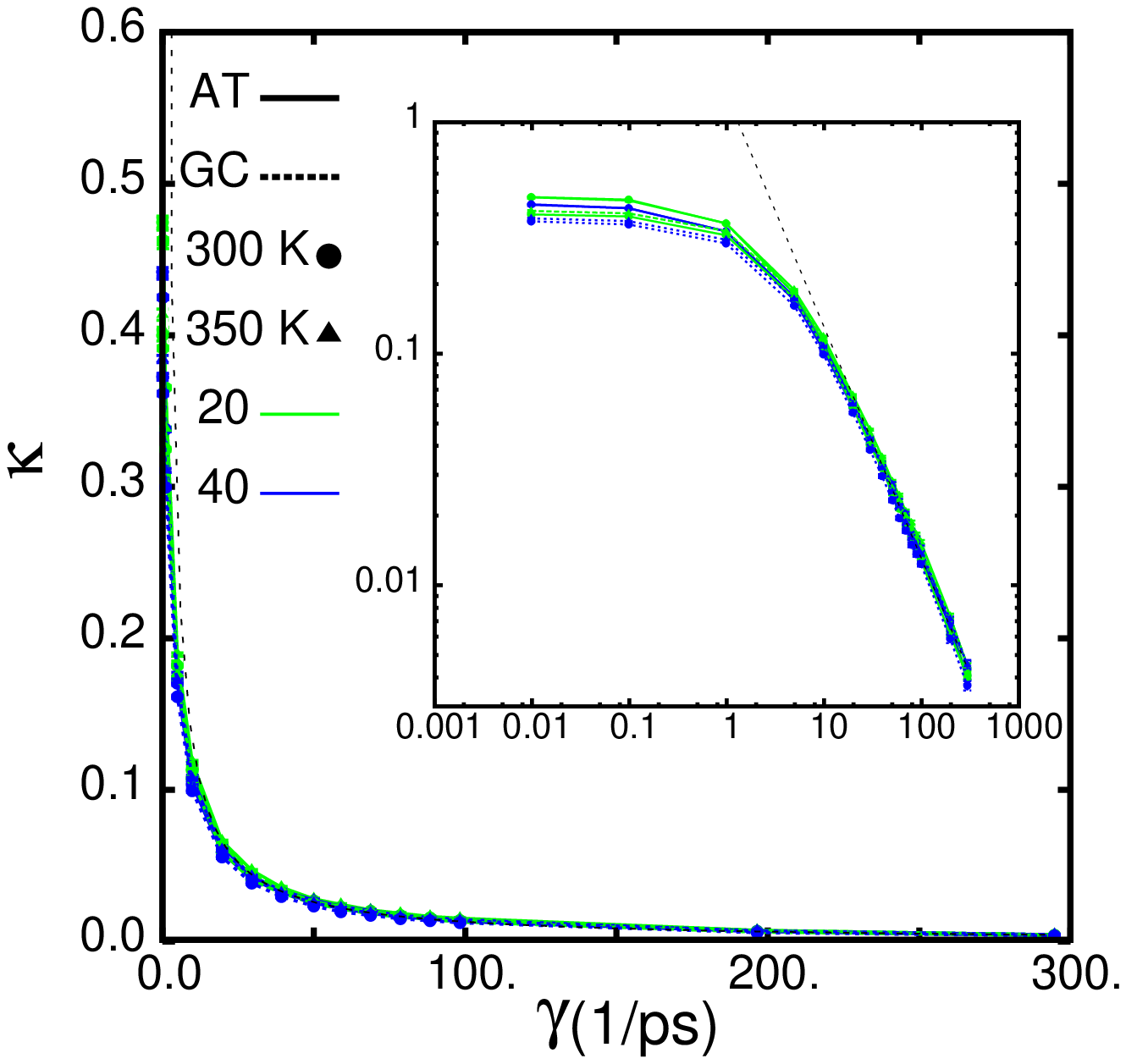}\\
   \caption{(color online) Transmission coefficient versus the friction coefficient $\gamma$ for a
homogeneous A- (solid line) or G- (dashed line) at temperatures $T=300$ K 
(spheres) or 350 K (triangles) consisting of 20 (green) or 40 (blue) bps.
Inset shows the same results in a log-log plot. Dashed black line corresponds to the fit $\kappa(\gamma)=1.3$ ps$^{-1}/\gamma$.
\label{figkappavsgamma}}
  \end{center}
\end{figure}

In Fig.~\ref{figkappavsgamma} we examined the influence of the friction constant $\gamma$.
As the dynamics do not change the equilibrium distribution
they can only have an effect on $\kappa$.
In agreement with previous finding, we see that the transmission coefficient 
is hardly affected by the polymer length,
the type of bps and  temperature. 
In conclusion,
$\kappa$ is mainly determined by the friction coefficient $\gamma$.  
At high friction we notice a standard Kramer's behavior~\cite{50Kramer}
$\kappa \propto 1/\gamma$  
which is somewhat surprising given the exotic nature of the RC that
depends on all $y_i$ coordinates in a discontinuous way.
Contrary to Kramer's theory, 
$\kappa$  converges to values in the range $0.38-0.48$ and  does not approach 1 in the limit
$\gamma \rightarrow 0$.  

\begin{figure}[ht!]
  \begin{center}
  \includegraphics[angle=-0,width=6.5cm]{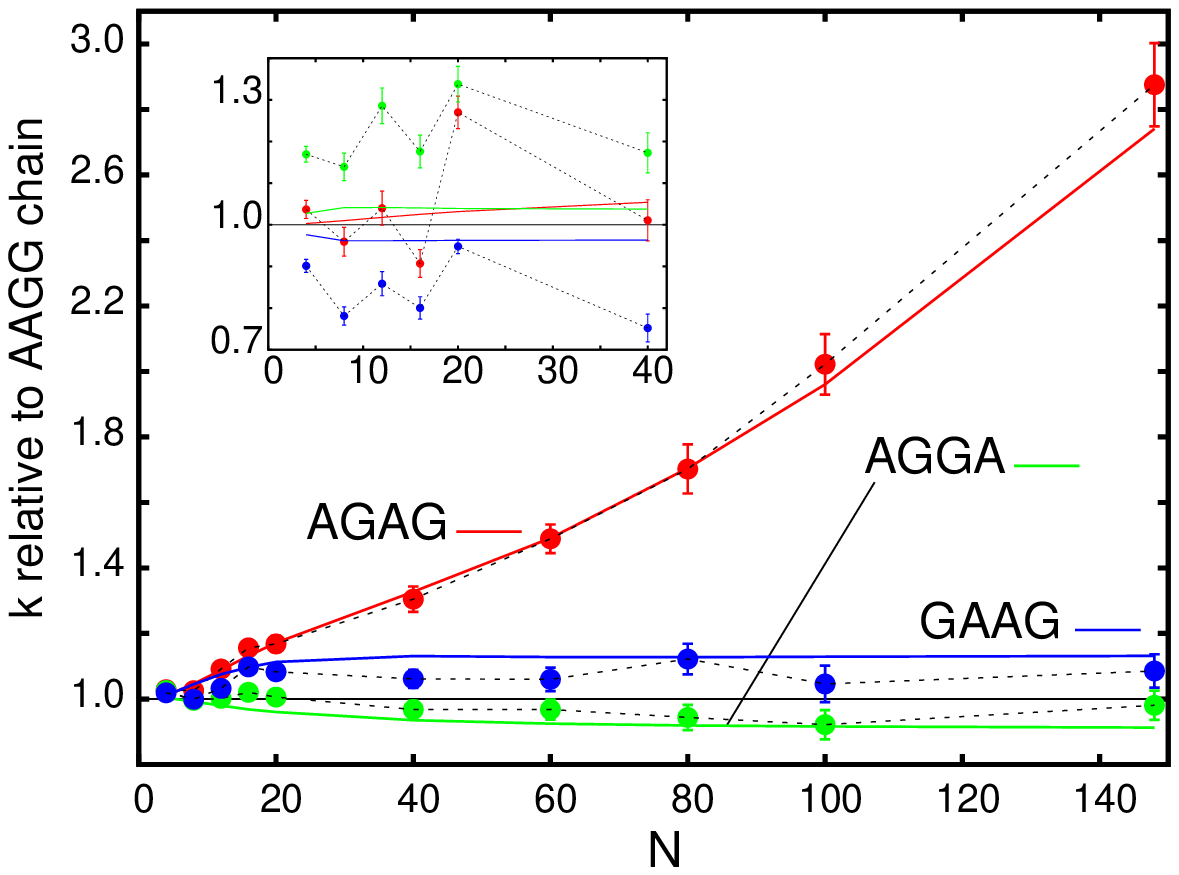}\\
   \caption{(color online) Relative denaturation rates versus chain length for 
DNA sequences having 50\% A- and 50\% G-bps. The curves only differ in the ordering of bps. All values are normalized by the chain having all A's at one side and the G's at the other side ("AAGG" after the N=4 bp chain). 
The alternating sequence AGAG (red) is shown together with the sequences 
having an A- (blue) or G-block (green) in the middle, respectively GAAG and
AGGA. Full circles with error-bars show the results of the full calculation 
while
the solid line is the result when transmission coefficients $\kappa$ would be considered to be identical for all sequences. Inset shows the same results 
for the alternative parameter set of Ref.~\cite{Theod}.
\label{figratio}}
  \end{center}
\end{figure}

Finally, we also examined the influence of the order of the sequence on the
denaturation rate. In fig.~\ref{figratio} we show
normalized denaturation rates as function of $N$ for sequences that all
contain 50\% A- and 50\% 
G-bases. 
The computed rate constant where  
normalized by the values of fig.~\ref{figkvsN2}. Hence,  all values 
are relative
to a chain having 
all A's and 
all G's at one side of the polymer. We call this the AAGG-sequence after the $N=4$ 
polymer having this characteristics. 
Besides the AAGG-sequence, we examined the alternating sequence AGAG, the
"weak-block in the middle" sequence GAAG, and the
"strong-block in the middle" sequence AGGA. 
The results were computed in two ways i) using the full rates, ii)
using the probabilities densities $P$ alone. The second approach basically 
assumes that all transmission coefficients are identical
for all sequences which seems to be right considering previous conclusions
and has the advantage that its result is not affected by statistical uncertainties. 

Fig.~\ref{figratio} shows
that 
the AGAG sequence 
gives significant faster denaturation compared to the AAGG-chain
and this effect 
increases with chain length. The GAAG-chain is second best being approximately 13\% faster than AAGG for $N>20$ without further increase 
as function of chain length. Conversely, the AGGA-sequence is about 10\%
slower than AAGG. Naturally, the bases at the end of the polymer are the most
easy to open. Apparently, the best way to profit from this effect is to put
the bases at the end that are difficult to open otherwise, but an even distribution of the G-bases
is by far the best.

The results of the alternative parameter set, however, 
(inset in fig.~\ref{figratio}) seems to give a somewhat different 
conclusion. The statistical error-bars are much larger 
for this parameter set due to the very weak coupling 
$K$ which increases  the accessible phase-space.
Still, AGGA and GAAG seem to be reversed regardless method and i) or ii) is applied. 
The AGGA seem to denature even faster than the 
AGAG-chain though this might be just   a statistical fluctuation.
If we consider the method ii), we see that the red curve starts 
to grow as function of $N$ and surpasses the AGGA result at $N=40$. 
Therefore, we believe that the alternating sequence has the fastest denaturation irrespective of 
the parameter set for sufficiently large $N$.
Besides their interest for studies involving real
  devices, those results on the effect of the sequence are important
  because they provide experimentally testable data \cite{Bonnet}
that discriminate
  between two parameter sets that give very similar equilibrium
  properties. This offers a unique possibility to validate model parameters.

In conclusion, we have utilized an innovative approach to study the denaturation
rate as function of different parameters. 
This has resulted in  several predictions which can be tested
experimentally. 
DNA models, giving similar results regarding statistics,
can differ fundamentally when denaturation  are concerned.
Therefore, our methodology in combination
with experiments provides an additional 
dimension to probe
the validity of DNA models and to improve them. 
This will be a significant step forward in understanding the 
sequence-dependent dynamics of DNA which play a  crucial role in 
several biological processes and new DNA-based devices
for  which the time response is important.

\acknowledgments
TSvE acknowledges 
the Methusalem funding (CASAS) 
by the Flemish government.

\bibliographystyle{prsty}

\end{document}